\documentclass[12pt]{article}

\usepackage{amssymb}
\usepackage{latexsym}

\usepackage{epsfig}
\usepackage{graphicx}

\usepackage{amsmath,amsthm,amsfonts,amssymb}

\newcommand{\be}{\begin{equation}}
\newcommand{\ee}{\end{equation}}
\newcommand{\bea}{\begin{eqnarray}}
\newcommand{\nn}{\nonumber}
\newcommand{\eea}{\end{eqnarray}}

\begin{document}

\begin{titlepage}
\begin{flushright}

gr-qc/xxxxxxx\\

\end{flushright}

\begin{centering}
\vspace{.41in}
{\large {\bf Braneworld Cosmological Models}}\\

\vspace{.4in}

 {\bf E.~Papantonopoulos} \\
\vspace{.2in}

 National Technical University of Athens, Physics
Department, Zografou Campus, GR 157 80, Athens, Greece. \\

\end{centering}
\vspace{1.3in}

\begin{abstract}

The main cosmological models on the brane are presented. A generic
equation is given, from which the Friedmann equations of the
Randall-Sundrum, induced gravity, Gauss-Bonnet and the combined
induced gravity and Gauss-Bonnet cosmological models are obtained.
We discuss the modifications they bring to the standard cosmology
and the main features of their inflationary dynamics.
\end{abstract}

\vspace{2.0in}
\begin{flushleft}

\uppercase{T}alk given at
 \uppercase{P}ascos04/\uppercase{N}ath \uppercase{F}est, \uppercase{B}oston,
\uppercase{A}ugust 16-22, 2004.\\
e-mail address:~lpapa@central.ntua.gr.

\end{flushleft}

\end{titlepage}

\section{Introduction}
The recent observation data from the Wilkinson Microwave
Anisotropy Probe (WMAP) \cite{wmap}, show strong support for the
standard inflationary predictions of a flat Universe with
adiabatic density perturbations in agreement with the simplest
class of inflationary models \cite{inflation}. In particular,
these observations had significantly narrowed the parameter space
of slow-roll inflationary models. We are entering an area where
the physics in the early universe can be probed by upcoming
high-precision observational data.

In light of these developments, it is important to understand
further the inflationary scenario from the theoretical point of
view and also from a more phenomenological approach. A new idea
that was put forward is, that our universe lies in a
three-dimensional brane within a higher-dimensional bulk spacetime
and this idea may have important consequences to our early time
Universe cosmology. The most successful model that incorporates
this idea is the Randall-Sundrum model of a single brane in an AdS
bulk \cite{randall}. There are also other brane cosmological
models which give novel features compared to standard cosmology.
These models are mainly generalizations of the Randall-Sundrum
model. The induced gravity cosmological model \cite{hc,dvali1}
arises when we add to the brane action, the generated
four-dimensional scalar curvature term by localized matter fields
on the brane. The Gauss-Bonnet cosmological model
\cite{dl,germani,charmousis,davis,gbmore} arises when we  include
a Gauss-Bonnet correction term to the five-dimensional action.
Finally, if both terms are included in the action, the combined
cosmological model \cite{papa} describes their cosmological
evolution.

In this talk we will give a brief account of these models,
describing the modifications they bring to the standard cosmology
and discussing their inflationary dynamics. We will follow a
general way of presentation writing the five-dimensional action
that includes a Gauss-Bonnet term in the bulk and a
four-dimensional scalar curvature term on the brane. Then, we will
derive a generic cubic equation in $H^{2}$, from which taking
appropriate limits of the parameters, we will derive the Friedmann
equations of the above cosmological models.

\section{Braneworld Cosmological Models}

Consider the five-dimensional gravitational action
 \bea
&& S_{\rm grav}=\frac{1}{2\kappa_{5}^{2}}\int
d^5x\sqrt{-^{(5)\!}g}
\left\{\,^{(5)\!}R-2\Lambda_{5}+\alpha\,\Big[\,^{(5)\!}R^{2}
\right.\nn
\\
&&\left.~{}-4\,^{(5)\!}R_{AB}\,^{(5)\!}
R^{AB}+^{(5)\!}R_{ABCD}\,^{(5)\!}R^{ABCD}\,\Big]\right\}\nn
\\
&&~{} + \frac{r}{2\kappa_{5}^{2}}\int_{y=0} d^4x\sqrt{-^{(4)\!}g}
\left[\,^{(4)\!}R-2\Lambda_{4}\right]\,, \label{action}
 \eea
where $\alpha$ is the Gauss-Bonnet coupling with dimensions
$(length)^{2}$ which is defined by
 \bea
\alpha=\frac{1}{ 8g_{s}^{2}}\,,
 \eea
with $g_{s}$ the string energy scale. The induced-gravity is
specified by the crossover length scale
 \bea
r=\frac{\kappa_5^2}{\kappa_4^2}=\frac{M_4^2}{M_5^3}\,.
 \label{distancescale}
 \eea
Here, the fundamental ($M_5$) and the four-dimensional ($M_4$)
Planck masses are given by
 \bea
\kappa_{5}^{2}=8\pi G_{5}=M_{5}^{-3}\,,~~ \kappa_{4}^{2}=8\pi
G_{4}=M_{4}^{-2}\,. \label{planck}
 \eea
We assume there are no sources in the bulk other than $\Lambda_5$.
We assume there are no sources in the bulk other than $\Lambda_5$.
Varying Eq.~(\ref{action}) with respect to the bulk metric
$^{(5)\!}g_{AB}$, we obtain the field equations:
 \bea
&& ^{(5)\!}G_{AB} -
\frac{\alpha}{2}\left[\,^{(5)\!}R^{2}-4\,^{(5)\!}R_{CD} \,
^{(5)\!}R^{CD}\right.\nn\\ && \left.~{}
+\,^{(5)\!}R_{CDEF}\, ^{(5)\!}R^{CDEF}\right]\,^{(5)\!}g_{AB}\nn \\
&&~{}+2\alpha\left[\, ^{(5)\!}R \,^{(5)\!}R_{AB}
-2\,^{(5)\!}R_{AC}\,^{(5)\!}R_{B}{}^C\right.\nn\\
&&\left.~{}-2\,^{(5)\!}R_{ACBD}\,^{(5)\!}R^{CD}
+\,^{(5)\!}R_{ACDE}\,^{(5)\!}R_{B}{}^{CDE}\right]\nn \\
&&{}= -\Lambda_{5}\,^{(5)\!}g_{AB}+\kappa_{5}^{2}\,^{\rm
(loc)}T_{AB}\hat{\delta}(y)\,, \label{varying}
 \eea
where $^{(4)\!}g_{AB}=\,^{(5)\!}g_{AB}-n_{A}n_{B}$ is the induced
metric on the hypersurfaces $\{y=$ constant\}, with $n^{A}$ the
normal vector. The localized energy-momentum tensor on the brane
is
 \bea
^{\rm (loc)}T_{AB} \equiv \,^{(4)\!}T_{AB}- \lambda
\,^{(4)\!}g_{AB}-\frac{r}{\kappa_{5}^{2}}\, ^{(4)\!}G_{AB}\,,
\label{tlocal}
 \eea where $\lambda $ is the brane tension and we have used the normalized Dirac delta function, $
\hat{\delta}(y)=\sqrt{^{(4)\!}g/\,^{(5)\!}g}\,\,\delta(y)$. Note
that the last term in (\ref{tlocal}) is due to the
 presence of the scalar curvature term on the brane.

 From the
 action (\ref{action}), for a homogeneous and isotropic brane at fixed coordinate position
$y=0$ in the bulk, we get the cubic equation in $H^{2}$,
\cite{papa} \bea &&{4\over r ^2}\left[1
+\frac{8}{3}\alpha\left(H^2 +{k \over a^2} + {\Phi_{0}\over 2}
\right) \right]^{2}\left(H^2 +{k \over a^2}-\Phi_{0}\right) \nn \\
&&~~~{} =\left[H^2 +{k \over a^2} -\frac{\kappa^{2}_{4}}
{3}(\rho+\lambda)\right]^{2}\,, \label{3fried}
 \eea
where $\Phi_0=\Phi(t,0)$ and $\Phi$ is a solution of the equation
$\Phi+2\alpha \Phi^{2}=\Lambda_{5}/6+\mathcal{C}/a^{4} $ with
$\mathcal{C}$ an integration constant, known as the `dark`
radiation term and in $\Lambda_{5}$ the coupling $\kappa^{2}_{5}$
has been absorbed.

The fundamental parameters appearing in (\ref{3fried}) are: three
energy scales, i.e. the fundamental Planck mass $M_{5}$, the
induced-gravity crossover energy scale $r^{-1}$, and the
Gauss-Bonnet coupling energy scale $\alpha^{-1/2}$, and two vacuum
energies, i.e. the bulk cosmological constant $\Lambda_{5}$ and
the brane tension $\lambda$. The parameters $r^{-1}$ and
$\alpha^{-1/2}$ are independent to each other. The crossover scale
$r$ of the induced gravity appears in loops involving matter
particles, and depending on the mass, it can be arbitrarily large.
On the other hand, the Gauss-Bonnet coupling $\alpha$ arises from
integrating out massive string modes, and depending on the scale
of the theory, it can also be arbitrarily large. From the generic
cubic equation (\ref{3fried}), taking the appropriate limits of
the above parameters, we will generate the Friedmann equations of
the various known cosmological models.

\subsection{ The
Randall-Sundrum Model}

Taking the limits $r\rightarrow 0$ and $\alpha\rightarrow 0$ in
(\ref{3fried}), we get the Friedmann equation of the
Randall-Sundrum model \cite{binetruy,csaki} \be
H^{2}+\frac{k}{a^{2}}=\frac{\kappa_{5}^{4}}{36}\rho^{2}+\frac{C}{a^{4}}+\frac{\kappa_{5}^{4}}{18}\lambda\rho
+\frac{\Lambda_{5}}{6}+\frac{\kappa_{5}^{4}}{36}\lambda^{2}
\label{randsund}. \ee In the early universe $\rho >> \lambda$, and
the $\rho^{2}$ term in (\ref{randsund}) is dominant, giving a high
energy modification of the standard Friedmann equation. To recover
the late time cosmology we define the four-dimensional Newton's
constant as $8 \pi G=\kappa_{5}^{4}\lambda /6$. Note that if
$\lambda=0$ we cannot recover the late time cosmology in this
model. There is also an effective cosmological constant given by
the last two terms of (\ref{randsund}) \be
\Lambda_{eff}=\frac{\Lambda_{5}}{6}+\frac{\kappa_{5}^{4}}{36}\lambda^{2}
\label{effconst}. \ee Hence in principle, it is easy to have an
accelerating phase in the Randall-Sundrum model. The second term
in (\ref{randsund}) is also a new brane-effect term called `dark`
radiation, because it scales as $a^{-4}$. The constant $C$ is
having the information of the bulk and it is proportional to the
mass of the black hole in the bulk.

If we assume that the early accelerating phase is governed by a
scalar field, we can set $\Lambda_{eff}$ of (\ref{effconst}) equal
to zero. Then, the  Randall-Sundrum Friedmann equation
(\ref{randsund}) becomes \be H^{2}=
\frac{\kappa^{2}_{4}\rho}{3}\Big{(}1+\frac{\rho}{2\lambda}\Big{)}.\label{indrudsud}
\ee Applying the inflationary formalism to this Friedmann equation
it was found, that the $\rho^{2}$ term acts as friction term which
damps the rolling of the scalar field, allowing steeper potentials
\cite{maartens,liddle,shinji}. It was also found that this damping
effect can bring the value of the inflaton field below the Planck
mass, giving a solution to one of the basic problems of the
chaotic inflationary scenario \cite{maartens}.

\subsection{The Induced Gravity Model}

Taking the limit $\alpha\rightarrow 0$ in (\ref{3fried}), we get
the Friedmann equation of the induced gravity cosmological model
~\cite{deffayet,astro,mmt,ktt,sahni} \bea && H^2+\frac{k}{a
^{2}}=\frac{\kappa_{4}^{2}}{3}(\rho + \lambda)
+\frac{{2}}{r^2}\nn\\&&~{} \pm \frac{1}{\sqrt{3}r}\left[{
4\kappa_{4}^{2}(\rho+\lambda) - 2\Lambda_{5}+{12\over r^2}-
\frac{12\mathcal{C}}{a ^{4}}}\right]^{1/2}.  \label{igr}
 \eea
Because of the square root, in the early universe the linear term
is dominant and therefore the induced gravity model in the high
energy limit gives a correction to standard Friedmann equation,
while at late times there are significant modifications to the
Friedmann equation, and in the limit $a\rightarrow \infty$ a
linearization of (\ref{igr}) gives again the conventional
cosmology with an effective cosmological constant
 \be \Lambda_{\rm
eff}=\kappa_{4}^{2}\lambda+\frac{6}{r^{2}}\pm
\frac{\sqrt{6}}{r^{2}}\sqrt{\left(2\kappa_{4}^{2}\lambda-
\Lambda_{5} \right)r^{2}+6}\,\,.\label{lambdaIG} \ee

The cosmological evolution described by the induced gravity model
is a four dimensional evolution at high energies/early times
followed by a five-dimensional and at low energies/late times
again four-dimensional evolution \cite{ktt}. Applying the
slow-roll inflationary formalism
\cite{zamarias,zhang,mariam,wands} setting $\Lambda_{eff}$ of
(\ref {lambdaIG}) equal to zero and redefining the high energy
four-dimensional Newton's constant
$\kappa_{4}^{2}=\kappa^{2}_{Planck}/\mu$, where $\mu$ is a small
number, it was found a better agreement of the chaotic
inflationary scenario with observational data, with the value of
the inflaton field well below the Planck mass \cite{zamarias}.

\subsection{The  Gauss-Bonnet Model}

Taking the limit $r\rightarrow 0$ in (\ref{3fried}), we get the
Friedmann equation of the Gauss-Bonnet cosmological model
~\cite{germani,charmousis}
 \bea
H^{2} +\frac{k}{a^{2} }= \frac{1}{8\alpha}\left(-2+\frac{64I
^{2}}{J}+J\right)\,, \label{friedgb}
 \eea
where the dimensionless quantities $I , J$ are given by
 \bea
&&\!I=\frac{1}{8}(1+4\alpha\Phi_{0})=\pm\frac{1}{8}
\left[1+\frac{4}{3}\alpha\Lambda_{5}+
 \frac{8\alpha \mathcal{C}}{a ^{4}}\right]^{1/2}\!,
 \label{phi1}\\
&&\! J=\!
 \left[
\frac{\kappa^{2}_{5}\sqrt{\alpha}}{\sqrt{2}} (\rho +\lambda) +
\sqrt{{\kappa^{4}_{5}\alpha \over 2} (\rho + \lambda)^{2}
 +(8I )^{3} } \right]^{\!2/3}\!.\label{phi11}
 \eea
The Gauss-Bonnet cosmological model like the Randall-Sundrum model
gives modifications to the standard cosmology in the early
universe. In the high energy limit of (\ref{friedgb}) the dominant
contribution to the Friedmann equation is proportional to
$\rho^{2/3}$. The Friedmann equation (\ref{friedgb}) can be
written in a simpler form \cite{lidnun} \be
H^{2}=\frac{1}{4\alpha}\Big{[}b^{1/3}cosh\Big{(}\frac{2x}{3}-1
\Big{)}\Big{]}\label{lids}, \ee where x is defined by \be
\sigma=\Big{(} \frac{2b}{\alpha\kappa^{4}_{5}} \Big{)}sinhx, \ee
and $b$ is a constant. Applying the inflationary dynamics to the
Friedmann equation (\ref{lids}), it was found that if the
inflation is driven by an exponential inflaton field, the
Gauss-Bonnet term allows the inflationary parameters to take
values closer to the recent observational data \cite{lidnun} and
generally the presence of the Gauss-Bonnet term softens the
Randall-Sundrum constraints on steep inflation
\cite{sami,toporensky}.

\subsection{The Combined Model}
If both  $r$ and $\alpha$ are non-zero, the single real solution
of the cubic equation (\ref{3fried}) which is compatible with the
previously considered limits, gives the Friedmann equation of the
combined cosmological model having both the induced and
Gauss-Bonnet terms \cite{papa} \bea && H^{2}+\frac{k}{a ^{2}}
=\frac{4-3\beta }{12\beta\alpha }
\nn\\&&~~~{}-\frac{2}{3\beta\alpha } \sqrt{P ^{2}-6 Q
}\,\cos\left(\Theta\pm \frac{\pi}{3}\right)\,, \label{mama}
 \eea where  \bea
P & = & 1+3\beta I\,,\label{parameterP}
\\  Q
&=&\beta\left[\frac{1}{4} +I+\frac{\kappa_{4}^{2}\alpha}{3}
(\rho+\lambda)\right],\label{parameterQ}\\
\Theta( P ,Q
)&=&\frac{1}{3}\arccos\left[\frac{2P^{3}+27Q^{2}-18PQ}
{2(P^{2}-6Q)^{3/2}}\right] \!, \label{omega} \eea and the constant
$\beta$ is given by \be \beta={ 256\alpha \over 9r^2}\,.
 \ee
All the solutions of the combined cosmological model are of finite
density, independently of the spatial curvature of the universe
and the equation of state. As we discussed, the Gauss-Bonnet model
on its own dominates at early times and does not remove the
infinite-density singularity, while the induced gravity model on
its own mostly affects the late-time evolution. However, the
combination of these terms produces an ``interaction" that is not
obviously the superposition of their separate effects. In general
terms, the early-universe behaviour is strongly modified by the
effective coupling of the 5D curvature to the matter. The late
cosmological evolution of the combined model follows the standard
cosmology, even for zero brane tension, with a positive Newton
constant for one of the two branches of the solutions and positive
cosmological constant.

It was showed in \cite{papa} that a radiation brane can, for some
parameter values, undergo accelerated expansion at and near the
minimal scale factor, independently of the spatial curvature of
the universe. When there is a black hole in the bulk, a subset of
these solutions has infinite acceleration at $a_{0}$, which
signals the ``birth'' of an accelerated universe at finite energy,
but with a curvature singularity.

The Friedmann equation (\ref{mama}) can be rewritten in a more
familial form
\begin{eqnarray} \label{GBFried}
H^{2}&+&\frac{k}{a^{2}}=\frac{64(4-3\beta)}{27\beta^{2}r^{2}}
-\frac{16\sqrt{2}\xi}{9\beta r} \nonumber \\
&\times&\sqrt{-4\kappa^{2}_{4}(\rho+\lambda)+\frac{3}{8}\beta\Lambda_{5}
+\frac{512}{9\beta^{2}r^{2}}\Big{(}1-\frac{3}{2}\beta+\frac{9}{64}\beta^{2}\Big{)}
+\frac{9}{4}\beta\frac{C}{a^{4}}}
\end{eqnarray} where
$\xi=\cos(\Theta\pm\frac{\pi}{3})$, and its detailed early
inflationary dynamics is under investigation \cite{leeper}.

\section{Conclusions}

We have presented the main cosmological models on the brane and
discussed their inflationary dynamics. All of them are giving
modifications to the standard Friedmann equation, mostly in the
high energy/early time regime. The terms that modify the standard
Friedmann equation reflect the fact, that these models are defined
in more that four dimensions and therefore any agreement with
future observational data will give not only information on the
early cosmological evolution of our universe, but also information
on the structure of spacetime.

So far there is no any evidence that these models describe
correctly the early time cosmological evolution and all recent
astronomical, astrophysical and cosmological data can be
described, to large extent, within the general relativity theory.
Nevertheless, pursuing these theoretical ideas may help us to
understand better the recent and future observational data, detect
possible inconsistences and hopefully find some agreement between
observational data and theoretical predictions.

\section*{Acknowlegements}

This talk deports the work done in collaboration with G. Kofinas,
R. Maartens and V. Zamarias and it is partially supported by the
Greek Education Ministry research programs "Hrakleitos" and
"Pythagoras".


\begin{thebibliography}{0}
\bibitem{wmap} D. N. Spergel {\it et al.}, Astrophys. J. Suppl.
{\bf148}, 175 (2003), astro-ph/0302209.

\bibitem{inflation} A. H. Guth, Phys. Rev. {\bf D23}, 347 (1981);
A. Albrecht and P. J. Steinhardt, Phys. Rev. Lett. {\bf 48}, 1220
(1982); A. D. Linde, Phys. Lett. {\bf 108B}, 389 (1982); A. D.
Linde, Phys. Lett. {\bf 129B}, 177 (1983); J. M. Baardeen, P. J.
Steinhardt and M. S. Turner, Phys. Rev. {D28}, 679 (1983).

\bibitem{randall} L. Randall and R. Sundrum,
 Phys. Rev. Lett. {\bf 83}, 3370 (1999),
 hep-th/9905221; Phys. Rev. Lett. {\bf83}, 4690 (1999), hep-th/9906064.

\bibitem{hc}
H. Collins and B. Holdom,  Phys. Rev. {\bf D62}, 105009 (2000),
hep-ph/0003173.


\bibitem{dvali1}
G. Dvali, G. Gabadadze and M. Porati, Phys. Lett. {\bf 485B}, 208
(2000), hep-th/0005016; G. Dvali and G. Gabadadze, Phys. Rev. {\bf
D63}, 065007 (2001), hep-th/0008054.


\bibitem{dl}
N. Deruelle and T. Dolezel, Phys. Rev. {\bf D62}, 103502 (2000),
gr-qc/0004021; B. Abdesselam and N. Mohammedi, Phys. Rev. {\bf
D65}, 084018 (2002), hep-th/0110143.

\bibitem{germani}
C. Germani and C. Sopuerta, Phys. Rev. Lett. {\bf 88}, 231101
(2002), hep-th/0202060.

\bibitem{charmousis}
C. Charmousis and J. Dufaux, Class. Quantum Grav. {\bf 19}, 4671
(2002), hep-th/0202107.

\bibitem{davis}
S. Davis, Phys. Rev. {\bf D67}, 024030 (2003), hep-th/0208205; E.
Gravannis and S. Willison, Phys. Lett. {\bf 562B}, 118 (2003),
hep-th/0209076; N.E. Mavromatos and J. Rizos, Phys. Rev. {\bf
D62}, 124004 (2000), hep-th/0008074; Y.M. Cho, I.P. Neupane and
P.S. Wesson, Nucl. Phys. {\bf B621}, 388 (2002), hep-th/0104227.



\bibitem{gbmore}
I. Low and A. Zee, Nucl. Phys. {\bf{B585}}, 395 (2000),
hep-th/0004124; J.E. Lidsey, S. Nojiri and S. Odintsov, JHEP {\bf
06}, 026 (2002), hep-th/0202198; P. Binetruy, C. Charmousis, S.C.
Davis and J-F. Dufaux, Phys. Lett. {\bf B544}, 183 (2002),
hep-th/0206089; J.P. Gregory and A. Padilla, hep-th/0304250; N.
Deruelle and M. Sasaki, gr-qc/0306032; C. Barcelo, C. Germani and
C.F. Sopuerta, gr-qc/0306072.

\bibitem{papa}
G. Kofinas, R. Maartens and E. Papantonopoulos, JHEP {\bf 0310},
066 (2003), hep-th/0307138.

\bibitem{binetruy}
P. Bin\'etruy, C. Deffayet, U. Ellwanger and D. Langlois,  Phys.
Lett. {\bf 477B}, 285 (2000), hep-th/9910219.

\bibitem{csaki}
C. Csaki, M. Graesser, C. Kolda and J. Terning,  Phys. Lett. {\bf
462B}, 34 (1999) [hep-ph/9906513]; J. Cline, C. Grojean and G.
Servant, Phys. Rev. Lett. {\bf 83}, 4245  (1999), hep-ph/9906523.

\bibitem{maartens} R. Maartens, D. Wands, B. Bassett and I. Heard,
Phys. Rev. {\bf D62}, 041301 (2000), hep-ph/9912464.

\bibitem{liddle} A. R. Liddle and A. J. Smith, astro-ph/0307017.

\bibitem{shinji} S. Tsujikawa and A. R. Liddle, astro-ph/0312162.

\bibitem{deffayet} C. Deffayet, Phys. Lett. {\bf
502B}, 199 (2001), hep-th/0010186.

\bibitem{astro}
Y. Shtanov, hep-th/0005193; S. Nojiri and S.D. Odintsov,  JHEP
{\bf 07}, 049 (2000), hep-th/0006232; C. Deffayet,  Phys. Lett.
{\bf 502B}, 199 (2001), hep-th/0010186; G. Kofinas,  JHEP {\bf
08}, 034 (2001), hep-th/0108013; N.J. Kim, H.W. Lee and Y.S.
Myung, it Phys. Lett. {\bf 504B}, 323 (2001), hep-th/0101091; C.
Deffayet, G. Dvali and G. Gabadadze,  Phys. Rev. {\bf D65}, 044023
(2002), astro-ph/0105068; C. Deffayet, S.J. Landau, J. Raux, M.
Zaldarriaga and P. Astier,  Phys. Rev. {\bf D66}, 024019 (2002),
astro-ph/0201164.

\bibitem{mmt}
K. Maeda, S. Mizuno and T. Torii,  Phys. Rev. {\bf D68}, 024033
(2003), gr-qc/0303039.

\bibitem{ktt}
E. Kiritsis, N. Tetradis and T.N. Tomaras, JHEP {\bf 03}, 019
(2002), hep-th/0202037.

\bibitem{sahni}
V. Sahni and Y. Shtanov, Int. J. Mod. Phys. {\bf{11}}, 1 (2002),
gr-qc/0205111; U. Alam and V. Sahni, astro-ph/0209443.

\bibitem{zamarias} E. Papantonopoulos and V. Zamarias,  JCAP {\bf 10},
001 (2004) gr-qc/0403090.

\bibitem{zhang} H. Zhang and R-G. Cai, hep-th/0403234.

\bibitem{mariam} M. Bouhmadi-Lopez, R. Maartens and D. Wands,
hep-th/0407162.

\bibitem{wands} M. Bouhmadi-Lopez and D. Wands, hep-th/0408061.

\bibitem{lidnun}
J. E. Lidsey and N. J. Nunes, Phys. Rev. {\bf{D67}}, 103510
(2003), astro-ph/0303168.

\bibitem{sami} S. Tsujikawa, M. Sami, R. Maartens,
astro-ph/0406078.

\bibitem{toporensky} M. Sami, N. Savchenko, A. Toporensky,
hep-th/0408140.

\bibitem{leeper} E. Leeper, R. Maartens, E. Papantonopoulos, V.
Zamarias, work in progress.



\end{thebibliography}
\end{document}